\documentclass[useAMS,referee]{biom}
\usepackage{graphicx}
\usepackage{epstopdf}
\usepackage{booktabs}
\usepackage{amsmath}
\usepackage{multirow}
\usepackage{longtable}
\def\bSig\mathbf{\Sigma}

\title[An integrated heterogeneous Poisson model for neuron functions]{An integrated heterogeneous Poisson model for neuron functions in hand movement during reaching and grasp}

\author{SHU-CHUAN CHEN$^{1,*}$\email{scchen@isu.edu},
LUNG-AN LI$^{2}$, and JIPING HE$^{3}$ \\
$^{1}$Department of Mathematics and Statistics, Idaho State University, Pocatello, ID 83201, USA \\
$^{2}$Institute of Statistical Science, Academia Sinica, Taipei, Taiwan\\
$^{3}$School of Biological and Health Systems Engineering, Arizona State University, Tempe, AZ 85289, USA.
}

\begin{document}


\date{{\it Received October} 2017. {\it Revised February} 2008.  {\it
Accepted March} 2008.}



\pagerange{\pageref{firstpage}--\pageref{lastpage}}
\volume{64}
\pubyear{2017}
\artmonth{January}


\doi{10.1111/j.1541-0420.2005.00454.x}


\label{firstpage}


\begin{abstract}
{To understand potential encoding mechanism of motor cortical neurons for control commands during reach-to-grasp movements, experiments to record neuronal activities from primary motor cortical regions have been conducted in many research laboratories (for example, \cite{Don2002}, \cite{Sch1988}). The most popular approach in neuroscience community is to fit the Analysis of Variance (ANOVA) model using the firing rates of individual neurons. In addition to consider neural firing counts but also temporal intervals, \cite{Chen2012} proposed to apply Analysis of Covariance (ANCOVA) model. Due to the nature of the data, in this paper we propose to apply an integrated method, called {\it heterogeneous Poisson regression model}, to categorize different neural activities. Three scenarios are discussed to show that the proposed heterogeneous Poisson regression model can overcome some disadvantages of the traditional Poisson regression model.}

\end{abstract}

%



\maketitle


%

\section{Introduction}
\label{s:intro}
Clinically, trauma or neurological diseases often result in loss or deterioration of motor functions for the patients. Developing a cortically controlled neuroprosthetic system for rehabilitation and recovery of muscle control for motor functions becomes more urgent and demanding for
these people to live independently with a higher quality of life. For such systems to be feasible it is critical to have knowledge on how
to translate neuronal activities in relevant areas of central nervous system into practical control commands for actual motor
behaviors.  For the purpose of advancing our knowledge on the potential encoding mechanism of motor cortical neurons for control
commands during reach-to-grasp movements, experiments to record neuronal activities have been conducted in many research
laboratories (\cite{Fu1995}, \cite{Sch1994}).
However there are arguments on how to analyze this
type of data due to the unclear nature of the neural code. Two most commonly used coding mechanisms are neuronal firing rates (frequency) and intra-spike intervals (time).

Most neurobiologists adopt the simple approach that frequency
(rates) should be used to characterize functions of motor neurons. There
are two different approaches to analyze these types of data:
discrete time approach and continuous time approach. In the case of
continuous time approach, it is to pool the spike times of the
trials and then employ a stochastic process, basically a counting
process (for example, Poisson Process model).  Either the probability
distributions of spike trains can be described through the
conditional intensity function of the process, or the distribution
of the interval time between two spikes can be extended from the
stochastic process of spike counts. The difficulty of these methods
is the numerical estimation of the parameters for modeling these
processes. However, the likelihood of conditional Intensity
functions could be handled by generalized linear models (\cite{McC1989}) and software as discussed in \cite{Bri1988}.
 From this point of view, \cite{C2006} proposed
 to analyze the firing rates using logistic generalized additive model (\cite{Has1990}) including interactions to tell the difference for firing or not. From  the likelihood, Bayesian estimation has been proposed by \cite{Bro2004} as well.

One may consider the neuron firing rates to evaluate the difference
between neuron functions under different experimental conditions. Some standard statistical methods, for example
 Analysis of Variance (ANOVA), are commonly applied to study the associations between firing
 rates and the observed motor behavior under specific experimental conditions.  Because of several disadvantages including the basic
 underline assumption of ANOVA such that the response variables should be continuous for
 normality, it leads to the loss of statistical power. \cite{Chen2012} proposed to consider neural firing counts and temporal intervals and apply ANCOVA model for the analysis.

On the other hand, to overcome the drawback of the counting process
model, \cite{Kas2001} proposed Inhomogeneous Markov Interval (IMI) Processes. They first
discretized spike times into small intervals such that in each interval
there is either one or no spike to convert the data into
a binary sequence. In this case, numerical integration and particle
filters are involved in the estimation algorithms. Their approach
took advantage of discrete time approach.

The classical approaches are mainly in discrete time model
framework. The time is discretized into consequent small time
intervals and transition density is proposed for modeling the
behavior from the current interval to the next interval by ensembling
all firing of neurons. Although statistical Markov property could
not be directly employed, the recursive algorithm could be utilized
for tracing the movements. For example, extraction algorithm (\cite{Sch2004}) and population vector algorithm
(\cite{Sch1994}) have been employed to study reaching or drawing movements (\cite{Mor1999}).

The objective of this paper is to propose a novel method of analyzing this type of experimental data,
which would handle pooled data from various task
conditions; while each individual neuron could respond differently
as task condition changes. The motivation of the proposed model,  basically from the counting process, and a
generalized linear model, Poisson regression model, is
computation simplicity for each task condition.  When pooling all
firing counts from all different conditions, a novel model called
heterogeneous Poisson regression models, involving three stages, is proposed.

The layout of the paper is as the follows.  The detailed experiment
and data description will be briefly described in section 2.  In
section 3, the statistical analysis methodology called heterogeneous
Poisson regression models will be proposed, and three stages for model
building will be explained.
Three scenarios for which the traditional regression approach do not work are discussed and documented in section 4. Finally, the conclusion is given in
the last section.

\section{Data Description}
\label{description}
To study how the central system controls hand orientation and movement, an experiment was conducted by the Neural Interface Design of the Biodesign Institute at Arizona State University. The experimental protocol is reviewed and approved by ASU Institutional Animal Care and Use Committee. The detailed experiment and data description are given in Fan's dissertation (\cite{Fan2006}) and also summarized in the paper of \cite{Chen2012}. The activities of the motor cortical neurons of the trained monkeys were recorded when reaching and grasping the left or right (two directions) target at various orientations ($45^{\circ}$,
$90^{\circ}$, and $135^{\circ}$). This experiment generated six target conditions based on movement direction and target orientation (denoted by left$45^{\circ}$/left$90^{\circ}$/left$135^{\circ}$/right$45^{\circ}$/ right$90^{\circ}$/ right$135^{\circ}$). There were total 913 neurons with 18 replicates under each of these six task conditions. That is, for each neuron, the activities were recorded in the total of 108 successful trials. The activities of 913 neurons were then analyzed in this paper.
For the primary data analysis, the data was arranged according to the sequence of events
on each target condition, then calculated the spike counts
in each sequence of the event.

\begin{figure}
\begin{center}
  \includegraphics*[width=0.75\textwidth]{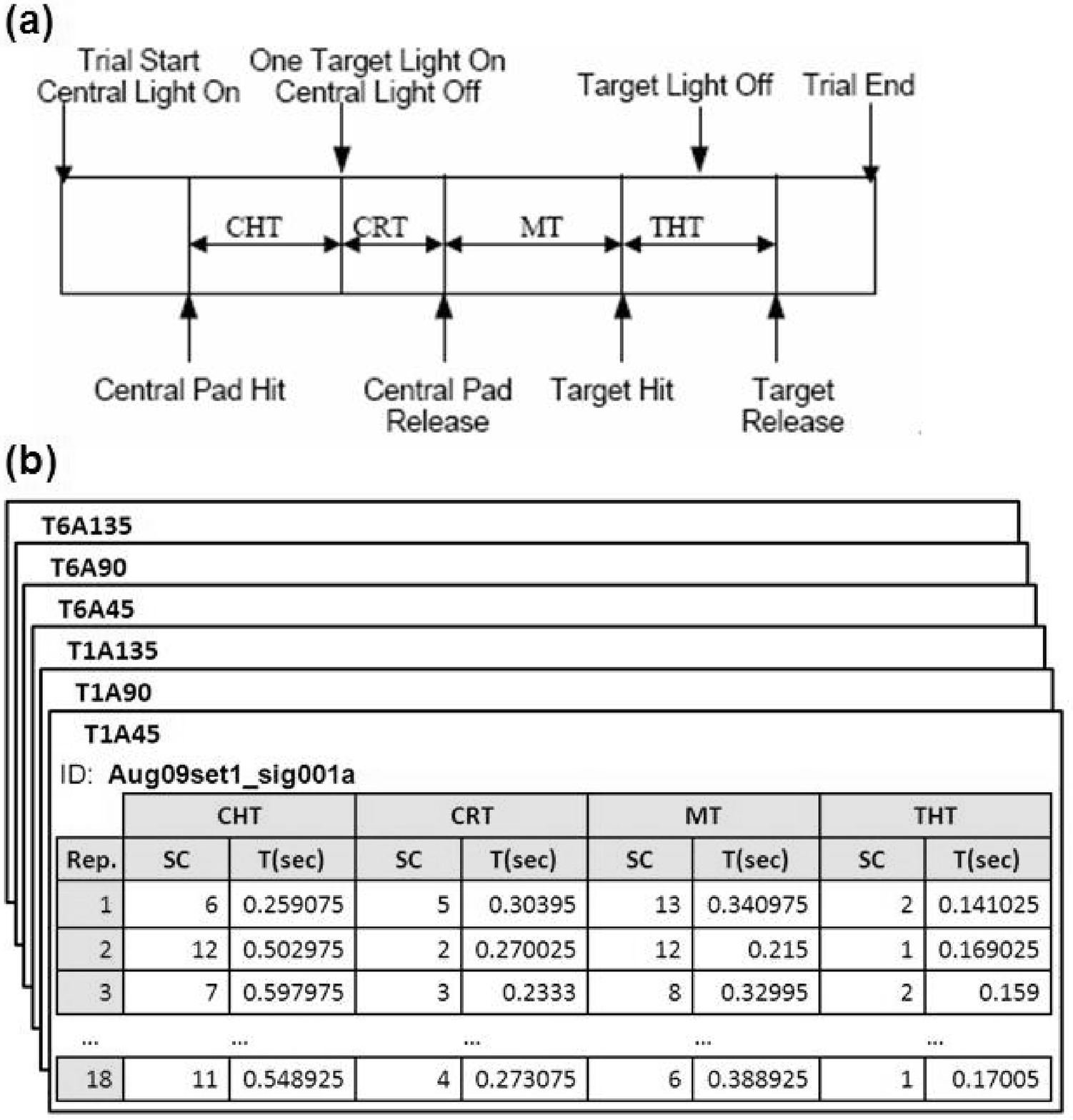}
\caption{(a)The sequence of events for the reach-to-grasp task are divided into four epochs: the center holding time (CHT) is for the time from central pad hit to target light on; the cue reaction time (CRT) is for the time from target light on to central pad release; the movement time (MT) is for the time from central pad release to target hit; the target holding time (THT) is for the time from target hit to target release. (b) Organized table of each neuron according to the target conditions, sequence of events. (SC-Spike Counts, T-Time duration of each epoch, direction left/right marked T1/T6, orientation $45^{\circ}$/$90^{\circ}$/$135^{\circ}$ marked A45/A90/A135)}
\label{fig:sequence of events}
\end{center}
\end{figure}

The recorded data of each neuron was organized as the tables shown in Figure~\ref{fig:sequence of events}(b).
Not all four epochs of these events are directly relevant to the
actual movement. Here we will focus on neuron activities during the time
from central pad release to target hit, marked as MT.


\section{Statistical Analysis Methodology}
\label{analysis}

If the neural spike counts have a linear trend
within MT duration, we expect the total firing counts of a
neuron over the time MT, $Y$, is a quadratic function of time MT.
\begin{equation}
\label{eq:MT}
    Y=\int^{t_0}_{0}0dt+\int^{MT}_{t_0}(a+bt)dt=\alpha_0+a*MT+\frac{b}{2}*MT^2,
\end{equation}
where ${t_0}$ is the starting time to fire in the MT duration. The Analysis of covariance (ANCOVA) approach was proposed in the paper of \cite{Chen2012} to analyze the data. However due to the nature of the data, the Poisson regression model might be better than ANCOVA while analyzing such type of count data. Therefore, in this paper, we propose {\it heterogeneous Poisson model} which consists of three stages that will be presented later.


\subsection{Data preprocess}
\label{sec:perprocess}

Graphic technique is a powerful tool as a complement to formal statistical methodology to capture the features of spike counts within MT duration time.
We used an ad hoc approach to visualize the fit of our statistical model and find potential outliers in the original data set.

We employed the LOWESS (locally weighted scatter plot smoothing) technique which was developed by \cite{W1979} on the log spike counts to
reflect the behavior of spike counts under each experimental condition.
Together with the fitted plot by the Poisson regression model, we could view if the spike counts did behave similarly as the pattern indicated by our model.
Spike counts are discrete, while LOWESS is a technique by smoothing the data within the neighborhood around the explanatory variable value to
obtain a smoothed response value through a polynomial function.
The polynomial is fitted using weighted least squares, giving more weight to points near the point whose response is being estimated and less weight to points further away.
By trial and error, we selected the window width to be 18 in our data.

From the plots of values from LOWESS and predicted values by Poisson regression under each target condition, we found most fits are quite reasonable for most neurons.
However, we noted a few exceptions, some outliers. We decided to remove those outliers before we fitted the model.
First we started by calculating the residuals from the local polynomial as
\begin{equation}
\label{eq:residual}
    e_i=Y_i-\hat{Y_i}
\end{equation}
where $Y_i$ is original spike count value and $\hat{Y_i}$ is the fitted value for local polynomial.

We classified the point as an outlier if the residual is $1.5\times IQR$ (the inter-quartile range) away from the LOWESS fitting.
That is, the inter-quartile range id defined by $IQR=Q_3-Q_1$, where $Q_1$ is the first quartile and $Q_3$ is the third quartile of those data within
the neighborhood around the value of the explanatory variable.
We used $1.5\times IQR$ to flag any observation which is either greater than $Q_3+1.5IQR$ or less than $Q_1-1.5IQR$ being considered as a suspicious point that could be called an outlier.
Next, the heterogeneous Poisson model will be presented.


\subsection{The First Stage: Elementary Models Construction}
\label{sec:Poisson} Poisson regression assumes that the response variable
$Y$ follows a Poisson distribution.  It assumes the logarithm of
its expected value can be modeled by a linear combination of unknown
parameters. In fact, the Poisson regression model is a special case of a
generalized linear model (GLM) with a log link.

For target conditions determined by directions (left, right),
and orientations ($45^{\circ}$, $90^{\circ}$, $135^{\circ}$), a
Poisson regression model was first fitted on 18 observations of each
target condition. The Poisson regression model is constructed as
\begin{equation}
\label{eq:poisson second}
    \log(E(Y))=\alpha+\beta X+\gamma X^2
\end{equation}
where $Y$ is the spike counts during the MT stage and $X$ represents the
MT duration  in microseconds.

\begin{figure}
\begin{center}
  \includegraphics*[width=0.75\textwidth]{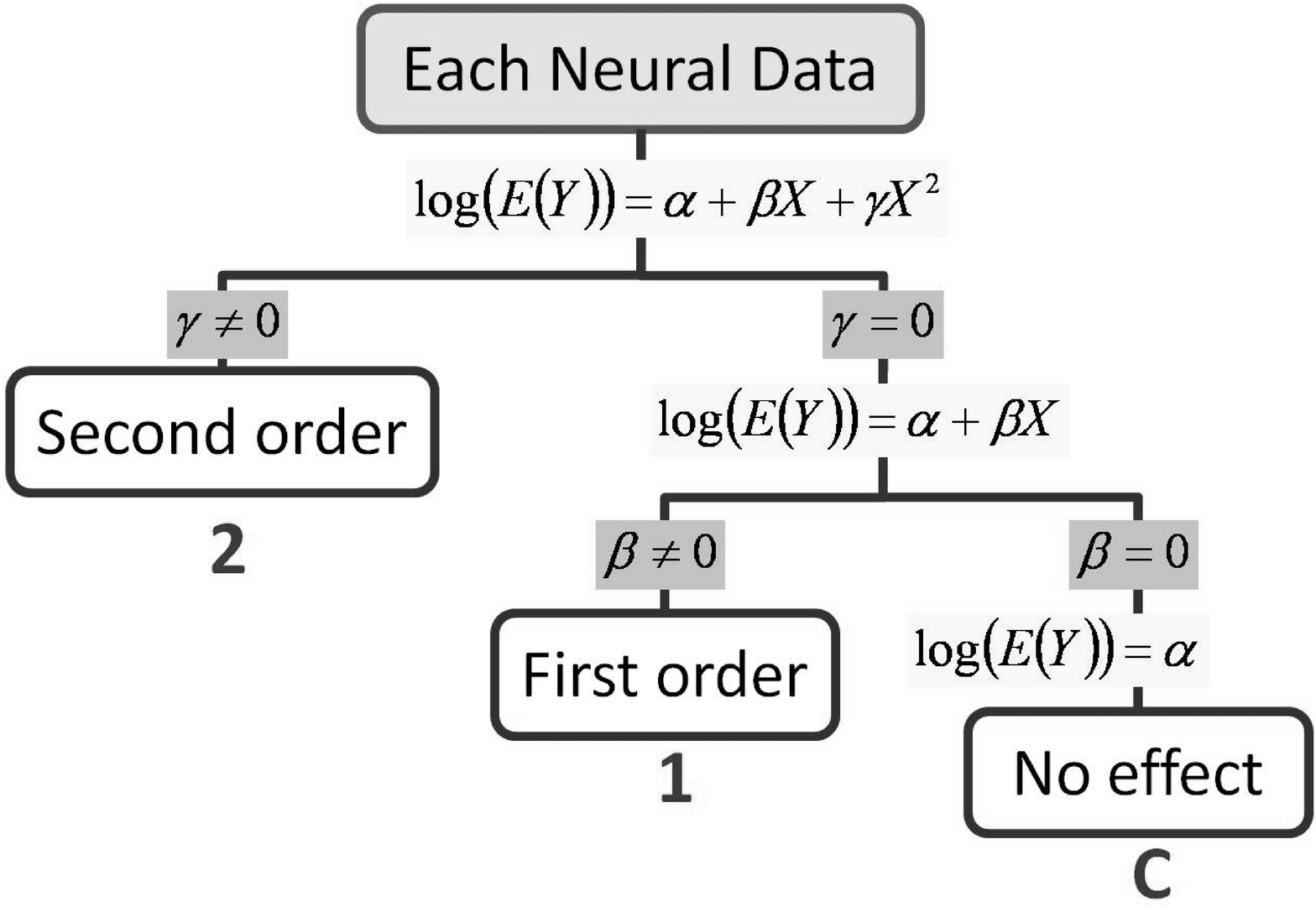}
\caption{The flow chart of Poisson regression models.}
\label{fig:flowChart}
\end{center}
\end{figure}

Figure~\ref{fig:flowChart} shows the sequence of actions within the process of the Poisson regression fitting for each target condition.

First, we check whether the neuron has the second order effect (whether the quadratic term of Poisson regression model is significant)
within different target conditions. We found that each neuron acted
differently at various target conditions. If the neuron has the
second order effect (the quadratic term of Poisson regression model
is significant), we mark the neuron with "2" to represent the second
order effect on that target condition. Here the sign of "2" or
"-2" represents the sign of the second order coefficient in the
model. If it is not of the second order effect, we then check the first order model.
If yes, we marked the neuron with "1" to represent the first order effect,
otherwise marked it with "C" to show no effect.
For all the neurons we were able to obtain the
elementary models for six target conditions. For example, Table 1 is for four neurons, Sep02set8\_sig004a, Sep15set3\_sig003a, Aug31set3\_sig005a, and Aug19set2\_sig002a.

\begin{table}
\begin{center}
\caption{Result of the Poisson regression model fitted on
observations for each target condition. Here 2 represents the second
order, 1 represents the first order, and C represents no effect.}
\label{tab:example}

\begin{tabular}{ccccccccc}
\hline\noalign{\smallskip}
\multicolumn{4}{p{5.5cm}}{(a) neuron: Sep02set8\_sig004a}   & & \multicolumn{4}{p{5.5cm}}{(b) neuron: Sep15set3\_sig003a}\\
\noalign{\smallskip}\hline\noalign{\smallskip}
Direction   & \multicolumn{3}{c}{Orientation}   &   & Direction & \multicolumn{3}{c}{Orientation} \\
\cmidrule[0.5pt](lr){2-4}\cmidrule[0.5pt](lr){7-9}
& A45   & A90   & A135  & & & A45   & A90   & A135  \\
\cmidrule[0.75pt]{1-9}
T1 & -1& C& 2& &T1 & -2& 2& 1\\
T6 & 2& -1& 2& &T6 & -2& 1& C\\
\hline \hline\noalign{\smallskip}
\multicolumn{4}{p{5.5cm}}{(c) neuron: Aug31set3\_sig005a}   & & \multicolumn{4}{p{5.5cm}}{(d) neuron: Aug19set2\_sig002a}\\
\noalign{\smallskip}\hline\noalign{\smallskip}
Direction   & \multicolumn{3}{c}{Orientation}   &   & Direction & \multicolumn{3}{c}{Orientation} \\
\cmidrule[0.5pt](lr){2-4}\cmidrule[0.5pt](lr){7-9}
& A45   & A90   & A135  & & & A45   & A90   & A135  \\
\cmidrule[0.75pt]{1-9}
T1 & -2& 1& C& &T1 &C &1 &1 \\
T6 & C& C& -2& &T6 &C &1 &C \\
\noalign{\smallskip}\hline
\end{tabular}
\end{center}
\end{table}

We also check if the plots of predicted values by Poisson model are similar to the plot of values from Lowess to diagnose the goodness-of-fitness of the model. For the example of neuron Aug17set1\_sig005a in the direction T6 and the orientation 45 degrees, it shows the plot of values from LOWESS is dissimilar to the pattern modeled by the second order Poisson regression model and test the effect of the quadratic term shows not significant. Hence, the fitted plot by the first order of Poisson regression model is similar to the plot of LOWESS, and test the linear term showed a significant effect. For all neurons, the fits are quite reasonable.


\subsection{The Second Stage: Initial Model Building}
\label{sec:perprocess} Based on those elementary models of six
target conditions described in previous subsection, we will next build the initial heterogeneous Poisson
model for pooling all observations from all target conditions. We will
explain this stage with an example described as follows. For example, the neuron
Sep02set8\_sig004a is shown in Table~\ref{tab:example}(a) to have
its elementary models respectively under six target conditions. In
which, the highest order among all the target conditions is the second
order effect. The initial heterogeneous Poisson model for this
neuron proposed in this paper is then described as
\begin{equation}
\label{eq:heterogeneous poisson model}
    \begin{aligned}
        \log(E(Y)) = &\alpha_1+\beta_1X+\gamma_1X^2\\
                &+D_{T1A45}(\alpha_2+\beta_2X+\gamma_2I_0X^2)\\
                &+D_{T1A90}(\alpha_3+\beta_3I_0X+\gamma_3I_0X^2)\\
                &+D_{T6A45}(\alpha_4+\beta_4X+\gamma_4X^2)\\
                &+D_{T6A90}(\alpha_5+\beta_5X+\gamma_5I_0X^2)\\
                &+D_{T6A135}(\alpha_6+\beta_6X+\gamma_6X^2)
    \end{aligned}
\end{equation}
where $Y$ is the spike counts during the MT stage and $X$ is the
MT duration in seconds for six target conditions respectively.
Here, $\gamma_4$ and $\gamma_6$ are coefficients for possible difference of the quadratic term
between baseline and other target conditions. The explanatory variables: $D_{T1A90}$, $D_{T1A135}$, $D_{T6A45}$, $D_{T6A90}$, and $D_{T6A135}$, take values on 0 and 1 that represent, except the base line setting condition,  other five target
categories' indicators. In this analysis, we especially add $I_0=-1$ for those target conditions that do not have the same efftect as  the baseline setting condition does. We will explain how we
handle these at the next subsection.


\subsection{The Third Stage: Final Model Establishment}
\label{sec:Heterogeneous} In this stage we use, once again, the
previous neuron as an example to demonstrate how to build the final
heterogeneous Poisson model. From the Poisson regression
fitting results of initial model, built in the previous stage on all
pooled observations of all six target conditions, we propose to
finalize the model by hypothesis testing. As we know some elementary
models do not have second order effect on those conditions, it
indicates that the neuron function at these target conditions might
be the first order effect or no effect when pooling observations. If
T1-with-orientation-45 and T6-with-orientation-90 conditions are
first order effect, we expect $\gamma_1=\gamma_2$ and
$\gamma_1=\gamma_5$ as $I_0=-1$ in these terms $\gamma_2I_0X^2$ and
$\gamma_5I_0X^2$ of (\ref{eq:heterogeneous poisson model}) in order
that the quadratic term of T1-with-orientation-45 and
T6-with-orientation-90 have zero coefficients. If
T1-with-orientation-90 condition has no effect, we expect
$\beta_1=\beta_3$ and $\gamma_1=\gamma_3$ as well. The results of
the fitting by the initial heterogeneous Poisson model for neuron
Sep02set8\_sig004a was shown in Supplementary material Table 1.

We propose to check the second order and first order effects by
considering the following testings:
\begin{equation}
\label{eq:test}
    \begin{aligned}
        &H_0:\gamma_1=\gamma_2 \mbox{ versus } H_a:\gamma_1\ne\gamma_2\\
        &H_0:\gamma_1=\gamma_3 \mbox{ versus } H_a:\gamma_1\ne\gamma_3\\
        &H_0:\gamma_1=\gamma_5 \mbox{ versus } H_a:\gamma_1\ne\gamma_5\\
        &H_0:\beta_1=\beta_3 \mbox{ versus } H_a:\beta_1\ne\beta_3\\
    \end{aligned}
\end{equation}
Under $H_0:\gamma_1=\gamma_2$ in (\ref{eq:test}), the deviance
equals to 1.387 and $df=1$. We believe zero coefficient of the
second order effect at T1-with-orientation-45 due to
$P-value=0.239$. For testing the second order of
T1-with-orientation-90 and T6-with-orientation-90, the deviances
equal to 3.179 ($P-value=0.074$) and 3.223 ($P-value=0.073$)
respectively, so it shows the second order effect at
T1-with-orientation-90 and T6-with-orientation-90 do not exist.
Taking $H_0:\beta_1=\beta_3$, the first order of
T1-with-orientation-90 has deviance equal to 0.879 with $df=1$, so
we might have no first order effect at T1-with-orientation-90
($P-value=0.348$).

\section{Why not traditional approach}

In this section, three scenarios that the traditional approach (for example, please refer to \cite{D1998}) does not work
for the analysis of neuron spike data will be presented. Motivating by the finding from these three scenarios, this leads to propose a novel model in the previous section.

\paragraph{Scenario 1: } Consider the baseline setting is the target condition with the highest order effect.

We considered neurons with at least one of elementary models being
of the second order effect. The set of neurons with at least one of elementary models being of
the second order effect has 444 neurons, 48.63\% of 913 neurons.
 What would it happen if we do not use the proposed model building approach in this paper, instead we followed the
 traditional approach. For demonstration, taking neuron Sep15set3\_sig003a  in
 Table~\ref{tab:example}(b) as an example, the baseline setting we suggested in our approach was chosen to be the
 target condition of direction T1 and orientation 45 degree, which is one of the highest
 order model among all target conditions. Then, a model for pooling all observations from all
 six target conditions was constructed as (\ref{eq:sc1}), (Please refer to p. 310 of \cite{D1998}) and the results after fitting this
  traditional regression model  are shown in Supplementary material Table 2. We easily see
  from P-values of those second order effect, it indicates that all target conditions are
   of the second order effect. However, this result of Supplementary material Table 2 is
   inconsistent with the result of Table~\ref{tab:example}(b).

\begin{equation}
\label{eq:sc1}
    \begin{aligned}
        \log(E(Y)) = &\alpha_1+\beta_1X+\gamma_1X^2\\
                &+D_{T1A90}(\alpha_2+\beta_2X+\gamma_2X^2)\\
                &+D_{T1A135}(\alpha_3+\beta_3X+\gamma_3X^2)\\
                &+D_{T6A45}(\alpha_4+\beta_4X+\gamma_4X^2)\\
                &+D_{T6A90}(\alpha_5+\beta_5X+\gamma_5X^2)\\
                &+D_{T6A135}(\alpha_6+\beta_6X+\gamma_6X^2)
    \end{aligned}
\end{equation}

In the set of neurons with at least one of elementary models being of
the second order effect, all 444 neurons have inconsistent problems with elementary models  if we fitted the traditional
approach. On the contrast, if by our proposed model in this paper,
except 25 neurons, 94.37 percent neurons are consistent with the elementary
models in this scenario. This scenario is taking one of target condition of the
second order effect as the baseline setting.

\paragraph{Scenario 2:} Consider the baseline setting is not any target condition with the
highest order effect among elementary models of six target
conditions.

In this scenario, we considered again the set of neurons with at
least one of elementary models being of the second order effect, but
we intentionally took one of the target condition not of the second
effect as the baseline setting for the model. We took neuron
Aug31set3\_sig005a in Table~\ref{tab:example}(c) as a demonstration
example. The baseline setting is the target condition with direction
T1 and orientation 90 degree, which is not the highest order among
all target conditions. The traditional model was constructed as
(\ref{eq:sc2-ex1all}) and the results of fitting this model shown in Supplementary Material
Table 3(a). Following the backward elimination procedure (Please refer to p. 239-342 of \cite{D1998}),
we sequentially removed non-significant
terms of the highest-order effect for each target condition. Doing
this by removing the quadratic terms of the largest P-value one
target condition by another target condition. After cleaning up
these second order effects, we checked the first-order terms for
each target condition too. The process stops if the remaining terms
are all significant. The final traditional approach model was shown
in (\ref{eq:sc2-ex1}) and the results of fitting the model was shown
in Supplementary material Table 3(b), however the first order effect of
baseline turned out to be not significant. This result is not
consistent with the result of Table~\ref{tab:example}(c) because the
model of the target condition with direction T1 and orientation 90
degree is of first order effect.

\begin{equation}
\label{eq:sc2-ex1all}
    \begin{aligned}
        \log(E(Y)) = &\alpha_1+\beta_1X\\
                &+D_{T1A45}(\alpha_2+\beta_2X+\gamma_2X^2)\\
                &+D_{T1A135}(\alpha_3+\beta_3X+\gamma_3X^2)\\
                &+D_{T6A45}(\alpha_4+\beta_4X+\gamma_4X^2)\\
                &+D_{T6A90}(\alpha_5+\beta_5X+\gamma_5X^2)\\
                &+D_{T6A135}(\alpha_6+\beta_6X+\gamma_6X^2)
    \end{aligned}
\end{equation}

\begin{equation}
\label{eq:sc2-ex1}
    \begin{aligned}
        \log(E(Y)) = &\alpha_1+\beta_1X\\
                &+D_{T1A45}(\alpha_2+\beta_2X+\gamma_2X^2)\\
                &+D_{T1A135}(\alpha_3)\\
                &+D_{T6A45}(\alpha_4)\\
                &+D_{T6A90}(\alpha_5)\\
                &+D_{T6A135}(\alpha_6+\beta_6X+\gamma_6X^2)
    \end{aligned}
\end{equation}

Among 444 neurons of this set we considered, 360 neurons, 81.08\%, are not
consistent with the elementary models by the traditional
 approach if the baseline setting is not of the second effect.

\paragraph{Scenario 3:} Consider the set of neurons with no target condition of the second order effect, but at least one target condition of the first order effect.
The set of neurons with no target condition of the second order effect, but at least one target condition of the first order effect has 419 neurons among 913 neurons.

For demonstration, taking neuron Aug19set2\_sig002a in
Table~\ref{tab:example}(d) as an example, the baseline setting was
the target condition of direction T1 and orientation 90 degree,
which is one of the highest order model among all target conditions.
Then, a model for pooling all observations from all six target
conditions was constructed as (\ref{eq:first}) and the results after
fitting this traditional regression model are shown in Supplementary material
Table 4. It indicates all target conditions are
of the first order effects. It is  inconsistent with the result of
Table~\ref{tab:example}(d).

\begin{equation}
\label{eq:first}
    \begin{aligned}
        \log(E(Y)) = &\alpha_1+\beta_1X\\
                &+D_{T1A45}(\alpha_2+\beta_2X)\\
                &+D_{T1A135}(\alpha_3+\beta_3X)\\
                &+D_{T6A45}(\alpha_4+\beta_4X)\\
                &+D_{T6A90}(\alpha_5+\beta_5X)\\
                &+D_{T6A135}(\alpha_6+\beta_6X)
    \end{aligned}
\end{equation}

All 419 neurons have inconsistent problems with  the elementary models by the traditional approach. On the contrary, if the proposed model in this paper is used, except 28 neurons, 93.32 percent neurons will be consistent with its elementary models in scenario 3. This scenario is taking one of target
conditions with the first order effect, which is the highest order effect as the baseline setting.


\section{Scientific analysis}
\label{sec:scientific analysis}


\subsection{Classification of neuron functions}
\label{class}

\begin{figure}
\begin{center}
  \includegraphics*[width=0.75\textwidth]{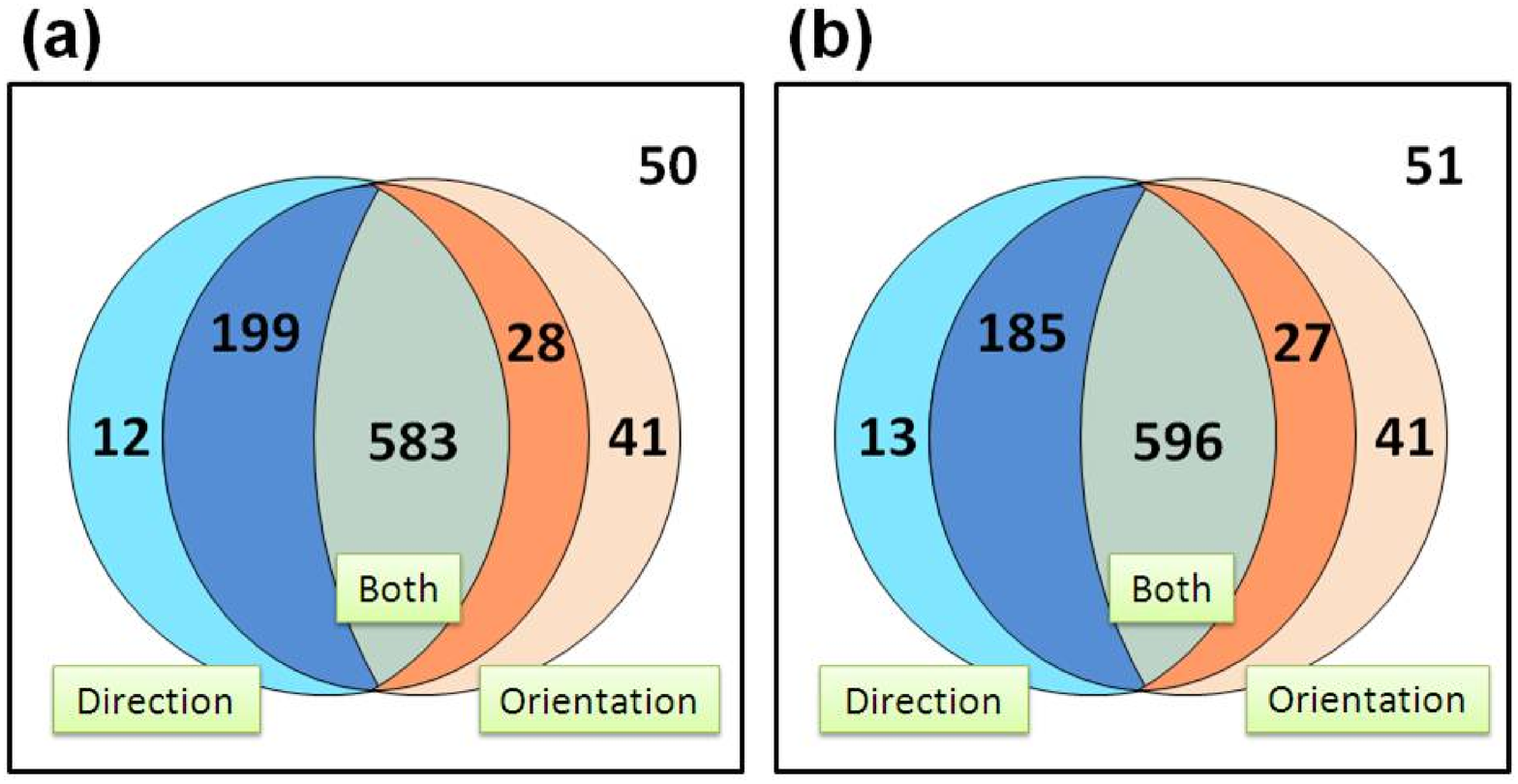}
\caption{Classification of 913 neurons. First, neuron's function is only related to direction (blue). Second, neuron's function is only related to orientation (orange). Third, neuron's function is related to specified direction and specified orientations (navy blue). Fourth, neuron's function is related to specified orientation but with favored direction (navy orange). Fifth, neuron's function is related both directions and orientations (gray). In addition, six target conditions of the neuron were all no effect (white frame). (a) Results of the Poisson regression model fittting. (b) Results of the heterogeneous Poisson model fitting.}
\label{fig:summary class}
\end{center}
\end{figure}

After fitting Poisson regression model, we learn neurons with significant second order effect or first order effect on some target conditions.
These information give us more insight on how the electric activities have been changed on different target conditions.
A systematic classification has been done for all of the recorded neurons. The 913 neurons were classified into six types, as shown in Figure~\ref{fig:summary class}(a). Based on the results obtained by the heterogeneous Poisson models for all neurons, we classify these 913 neurons and the results are shown in Figure~\ref{fig:summary class}(b). The results will be discussed next.

\subsubsection{Direction-related-only neurons}
\label{sec:Direction}

\begin{figure}
\begin{center}
  \includegraphics*[width=0.75\textwidth]{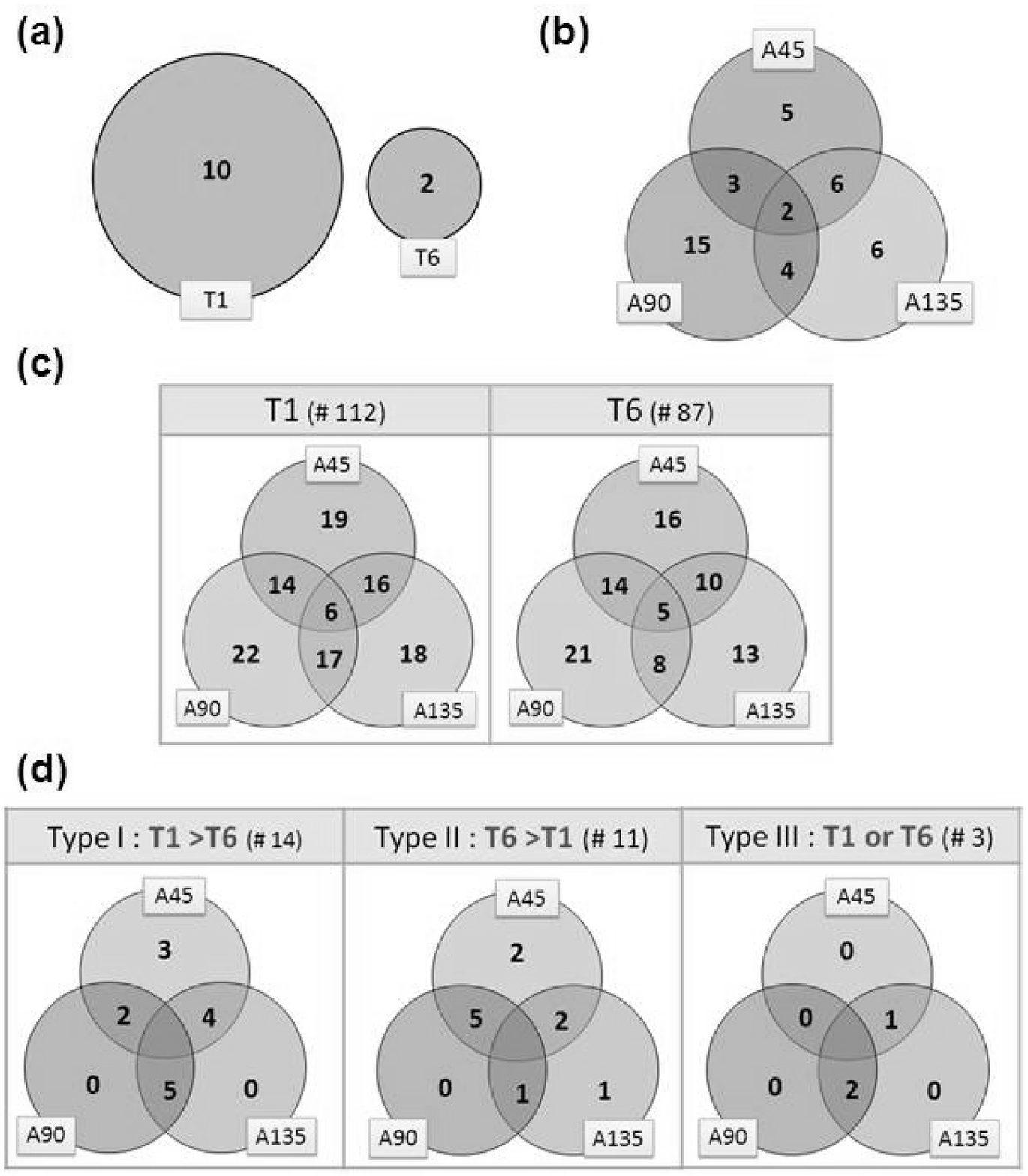}
\caption{Classification of neurons fitted the Poisson regression models. (a) Neuron's function is only related to direction. (b) Neuron's function is only related to orientation. (c) Neuron's function is related to specified direction and specified orientaions. (d) Neuron's function is related to specified orientation but with favored direction.}
\label{fig:class poisson}       
\end{center}
\end{figure}

\begin{figure}
\begin{center}
  \includegraphics*[width=0.75\textwidth]{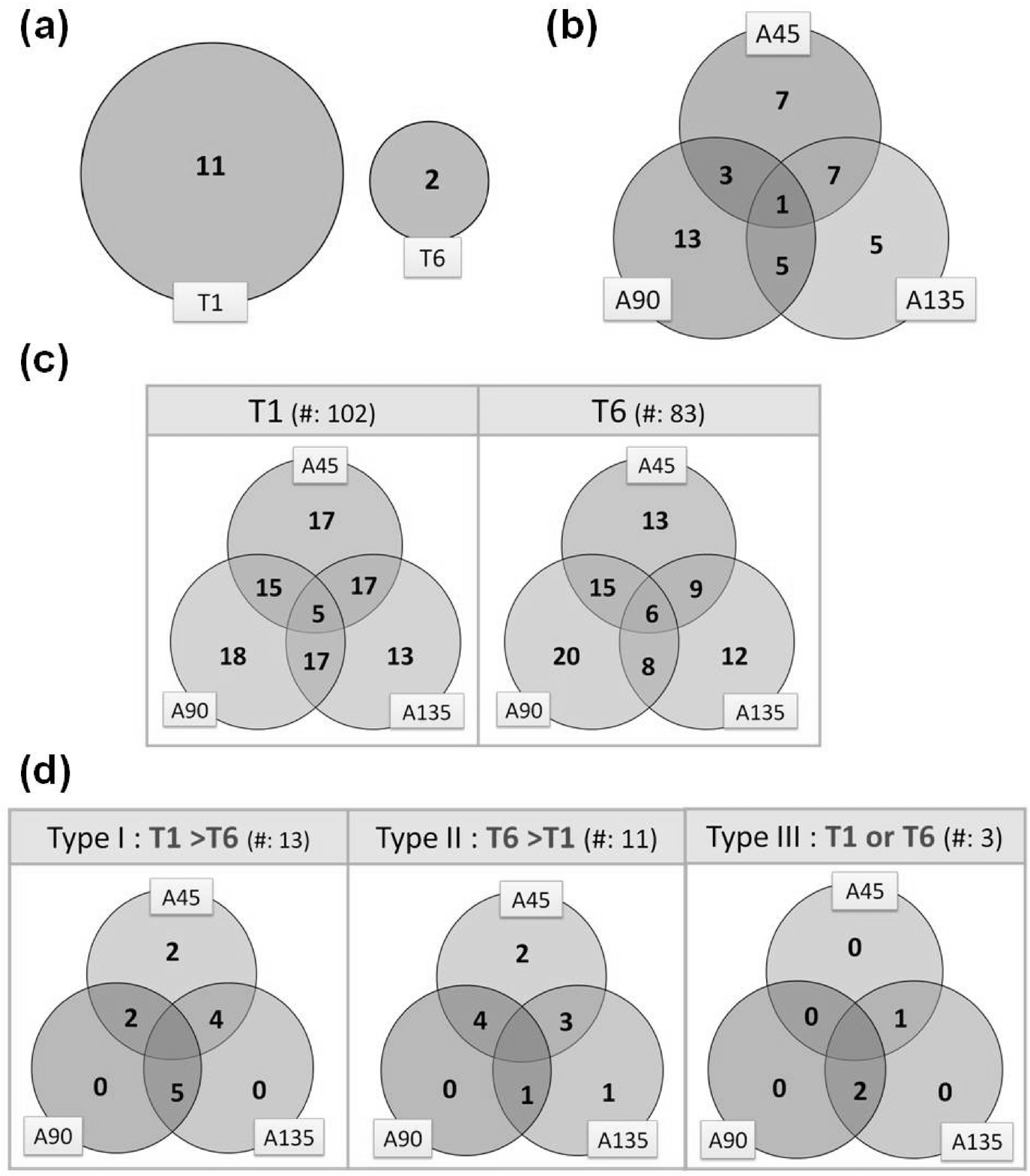}
\caption{Classification of neurons fitted the heterogeneous Poisson models. (a) Neuron's function is only related to direction. (b) Neuron's function is only related to orientation. (c) Neuron's function is related to specified direction and specified orientations. (d) Neuron's function is related to specified orientation but with favored direction.}
\label{fig:classHete}       
\end{center}
\end{figure}

In Table~\ref{tab:direction}(a), it shows two examples of neurons whose functions are only of the first order effects in direction T1 regardless of orientations, but no effect in direction T6. We classify these neurons as being only related to direction T1. In Table~\ref{tab:direction}(b), it's a case showing the first order effect in direction T6 regardless of orientations but no effect in direction T1.
These neurons were classified as being only related to direction T6. There are 12 neurons whose functions were classified as direction-related-only. The results are shown in Figure~\ref{fig:class poisson}(a).
They all are of the first order effect in directions, either in T1 or T6 regardless of orientations.
And Figure~\ref{fig:classHete}(a) shows classification of the heterogeneous Poisson model fitting.

\begin{table}
\begin{center}
\caption{Table of direction-related-only}
\label{tab:direction}       
\begin{tabular}{ccccccccc}
\hline\noalign{\smallskip}
\multicolumn{4}{c}{(a) Related to T1 direction}  & & \multicolumn{4}{c}{(b) Related to T6 direction}\\
\noalign{\smallskip}\hline\noalign{\smallskip}
Direction   & \multicolumn{3}{c}{Orientation}   &   & Direction & \multicolumn{3}{c}{Orientation} \\
\cmidrule[0.5pt](lr){2-4}\cmidrule[0.5pt](lr){7-9}
& A45   & A90   & A135  & & & A45   & A90   & A135  \\
\cmidrule[0.75pt]{1-9}
T1 & 1& 1& 1& &T1 & C& C& C\\
T6 & C& C& C& &T6 & 1& 1& 1\\
\noalign{\smallskip}\hline
\end{tabular}
\end{center}
\end{table}

\subsubsection{Orientation-related-only neurons}
\label{sec:Orientation}

Supplementary material Table 5(a) shows cases that neurons are of the first order effects in the orientation 45 degrees regardless of directions and
no effect in other conditions. We classify these neurons whose function are only related to orientation 45 degrees.
Similarly, those neurons whose functions are of the second order effects in orientation 45 and 90 degrees regardless of directions are shown in Supplementary material Table 5(b). We classify these neurons as being only related to orientation 45 and 90 degrees.
For those neurons their functions are uniformly related between the two directions, we classify them as being only related to orientation as in Supplementary material Table 5(c) or Table 5(d).
In total, 41 neurons are classified as orientation-related-only. The results are shown in Figure~\ref{fig:class poisson}(b). In Figure 7(b), it shows the classification results of the heterogeneous Poisson model fitting.


\subsubsection{Related to specified direction and specified orientation neurons}
\label{sec:specified direction specified orientation}

For direction T1 and the orientation 45 degrees in Supplementary material Table 8(a), they are of first order effect but no effect in other conditions,
so we classify these neuron as being related to T1-and-orientation-45.
Supplementary material Table 8(b) shows the first order effects in orientation 45 and 135 degrees given
direction T6. Given the specified direction, there are effects for this target condition where first order or second order in specified orientations as
shown in Supplementary material Table 8(c), and we also classify those neurons as being related to direction T1 and the orientation 45, 90, and135 degrees.
In total, 199 neurons are classified as related to specified direction and specified orientation. Figure~\ref{fig:class poisson}(c) shows the number of neurons on certain orientations and specified directions. In Figure~\ref{fig:classHete}(c), it shows the classification results of the heterogeneous Poisson model fitting.


\subsubsection{Related to specified orientation but with favored direction neurons}
\label{sec:specified orientation favored direction}

For some neurons, the orientation effect was not uniform between the two directions, so we classify these phonemon on three aspects.
First, there are significant effects in certain orientations, but direction T1 is of second order effect as shown in Supplementary material Table 9(a).
We classify these neuron as related to orientation 45 and 135 degrees but with favored direction T1.
Second, there are significant effects in certain orientations, but direction T6 is second order effect as in Supplementary material Table 7(b).
We classify these neurons as being related to orientation 45 degrees but with favored direction T6.
Third, the direction is either of second order effect or first order effect in different orientations as in Supplementary material Table 7(c),
we classify those neurons as being related to orientation 90 and 135 degrees but with either direction.
Therefore, there are 28 neurons classed as related to specified orientation but with favored direction.
Figure~\ref{fig:class poisson}(d) shows the number of neurons on certain orientations with favored direction respectively.
And Figure~\ref{fig:classHete}(d) shows classification of the heterogeneous Poisson model fitting.


\subsubsection{Direction-orientation-both-related neurons}
\label{sec:Direction orientation both}

Supplementary material Table 8(a) shows the significant effects in orientation-45-with-direction-T1 and orientation-90-with-direction-T6.
If the significant effects are first order or second order in two target conditions as shown in Supplementary material Table 8(b),
we classify these neurons as being orientation-specific and direction-specific.
There are 79 neurons classed as orientation-specific and direction-specific, and the number of neurons on certain two target conditions is given in
Supplementary material Table 9(a).
And Table 9(b) shows classification of the heterogeneous Poisson model fitting.

As for the remaining neurons, there are 504 neurons with different order effects and they may be considered as having interaction between direction and orientation.
In these cases, we utilize a simple counting method to calculate weights on orientation or direction.
Here we will show how we calculate weights. We give the score "2" to the second order effect, the score "1" to the first order effect, and the score "0" to no effect.
For each direction, the direction weight is the standardized sum of the scores across the orientations. Similarly, for each orientation, the orientation
weight is the standardized sum of the scores across the directions.
Here the standardization is done by its maximum possible score: the maximum direction possible score sum is 6 and the maximum orientation possible score sum is 4.
The proposed weights are given by (\ref{eq:weight}). Once the weights have been found, we choose the largest weight as the priority of
the favored specified direction or favored specified orientation.

\begin{equation}
\label{eq:weight}
\small
    \begin{aligned}
        \mbox{Direction weight}=& \frac{\mbox{Under certain direction, the sum of three orientation order effects}}{6}\\
        \mbox{Orientation weight}=& \frac{\mbox{Under certain orientation, the sum of two direction order effects}}{4}
    \end{aligned}
\end{equation}

For example in Supplementary material Table 10(a), direction T1 and the orientation 135 degrees have the largest weights.
Besides, the weight of the orientation 135 degrees is larger than direction T1.
We classify these neurons as being interaction between direction and orientation but more favor in orientation.
There are 294 neurons classed with more favor in orientation.
Table 10(b) shows those neurons with larger weights on direction than orientation, and then
we classify these neurons as interaction between direction and orientation but more favor in direction.
There are 77 neurons classed with more favor in direction.
If equal weights on orientation and direction as in Supplementary material Table 10(c), we classify these neurons as interaction between
direction and orientation but equally favor on direction or orientation.
There are 133 neurons classed with equally favor on direction or orientation.
Based on the results obtained by the heterogeneous Poisson models for 518 neurons with interaction between direction and orientation,
we classify them as Supplementary material Table 11.


\subsubsection{Neuron position}
\label{neuron position}
Among 913 neurons of this set we considered, 869 neurons, are recorded neuron position as Supplementary material Figure 1 and Figure 2.
For six types of classification, the summary of neuron position is shown in Supplementary material Table 12.

\section{Discussion and Conclusion}

For the heterogeneous Poisson model we proposed in this paper, we
strongly suggest to use the target condition of the highest order
effect among all six target conditions to be the baseline setting
for our model. If none of all six target conditions is of the second
order effect, we suggest using one of the target conditions being of the first
order effect in the initial model building.

Since in the last stage of our model building, we utilize hypothesis
testing procedure to find the final model, we recommend to adjust
the significance level for multiple tests.

In the first stage of our model building, the elementary model for
each target condition we built is the Poisson regression model on
observations for that target condition. We believe that model
reflects or gives a clue to the true model of that particular neuron
 under the environment of that target condition. Those
observations were independent under each target condition
environment for that particular neuron.

As the experiment was conducted independently over various target
conditions for the same neuron,we admitted we do not know how to
model or estimate the dependence structure relationship between
observations of different target conditions of the same neuron in this paper. However,
we found the model, the heterogeneous Poisson model proposed in this
paper works consistently with elementary models when pooling data
across various environments. There are only 53 neurons among 913
neurons we analyzed whose final model changed by our approach
compared to its  elementary model. We listed them in the appendix. In
conclusion, we proposed the heterogeneous Poisson model to analyze
the function of central nervous system when pooling various
environments.


\section*{Acknowledgements}

We thank our research assistants, Ms.  Wan-Ting Chen and Mr. Tun-Hao Chang for their assistance in this project. Dr. Chen also acknowledge the support from National Center for Theoretical Sciences (South), Taiwan.

\setcounter{table}{0}

\end{document}